# Estimation of Power Balance in Steady State LHCD Discharges on TRIAM-1M

T. Sugata[a], K. Hanada[b], N. Imamura[a], M. Sakamoto[b], H. Zushi[b], H. Idei[b], A. Iyomasa[b],
S. Kawasaki[b], K. Sato[b], H. Nakashima[b], K. Nakamura[b], M. Hasegawa[b], A. Higashijima[b]

[a]*Interdisciplinary Graduate School of Engineering Sciences, Kyushu University, Japan*
[b]*Research Institute for Applied Mechanics, Kyushu University, Japan*

**Abstract**

On TRIAM-1M, a long duration discharge for more than 5 h was achieved successfully by fully non-inductive lower hybrid current drive. Heat load distribution to the plasma facing components (PFCs) in the 5 h discharge was investigated by using calorimetric measurements. The injected RF power was coincident with the total amount of heat load to PFCs estimated by calorimetric measurement. The power balance including the portion of direct loss power of fast electron, heat flux due to the charge exchange process was able to be estimated in this long duration discharge.

**Keywords:**
Tokamak, Calorimetry, Power Balance, LHCD, TRIAM-1M

## 1. Introduction

The steady state operation is important for realize the fusion power plant. In ITER, over 1000 sec duration discharge has been planned [1]. In this case, the heat flux to the divertor will be estimated 5 MW/m$^2$ [2] and treatment of this local huge heat load is a key issue for steady state operation. One method for treatment of heat is effective cooling for plasma facing components (PFCs). In Tore Supra, the CIEL project using the toroidal pump limiter (TPL) has been enforced. TPL was designed to extract power up to 15 MW [3]. Exhausting huge heat load allows implementation of the injection of high power to plasmas. In fact, the CIEL project succeeded to maintain the discharge for 6 min. Total injected energy reached up to 1 GJ, which means that the averaged heat load of 2.8 MW could be removed from PFCs continuously. Although the total cooling capability was enough, the discharge was terminated contrary to the will. This suggests that the heat and particle handling, that is the control of the distribution of heat load, is not sufficient to maintain the plasma in steady state. The avoidance of the concentration of the heat load should be investigated. In fact, for ITER scenario a radiative power fraction of $\approx 75\%$ will be required to avoid the concentration of heat load to the divertor plate [4]. Therefore, the study of heat load distribution in steady state is very important. The distribution of heat load has been decided by the power balance of the plasma. When the distribution of heat load would be controlled, the knowledge of the power balance of the plasma plays an essential role. Therefore the study of the power balance should be executed.

On TRIAM-1M, it was obtained that more than 5 h discharge maintained by lower hybrid current drive (LHCD) with the quite low power region. This ultra long duration discharge is suitable to measure the heat load, because the water temperature is saturated during the discharge. As in this situation, the heat load can be derived from the steady temperature rise as shown in section 3, accuracy of estimation of the heat load become to be better. In this paper, the result of the heat load to each PFC is described and the power balance of the discharge is shown. In section 2 experimental apparatus is introduced and the way to measure the heat load is shown in section3. The experimental results are presented in section 4. The conclusion of this paper is summarized in section 5.

## 2. Experimental apparatus

TRIAM-1M is a small size ($R_0$ = 0.8 m, a×b = 0.12 m×0.18 m) tokamak with the high toroidal magnetic field up to 8 T excited by 16 toroidal field coils made of the super-conducting material, $Nb_3Sn$ [5]. TRIAM-1M has two vacuum vessels. One is to avoid thermal penetration into superconducting coils from outward. The vacuum level should maintain a low pressure level less than $1 \times 10^{-5}$ Pa. The other is to keep the vacuum condition of $1 \times 10^{-6}$ Pa or less for making plasma. Fig.1 is toroidal cross section diagrams of the latter vacuum vessel made of stainless steel and divertor plates. Numbering on the vacuum vessel shows the branch of the cooling water channel as described below. The shadow area on Fig.1 shows the region covered by divertor plates made of molybdenum, which have been installed on the down part of the vacuum vessel. As shown in Fig.1, five limiters covered by molybdenum are installed on the vacuum vessel. Three poloidal ring limiters are illustrated in Fig.1. These limiters are distributed in the toroidal direction to avoid the direct contact with the plasma to the vacuum vessel. Their size is the same value in width of 32 mm and in poloidal length of 1040 mm. These limiters are installed on stainless steel bases, which play a role in attachment of the fixed limiter on the surface of the vacuum vessel. These limiters could be cooled indirectly through water channels made of copper brazed on the stainless steel base. Especially, as the inner and outer part of limiters could not equip any water cooling channels, thermal transport of their parts is quite poor than the other part of fixed

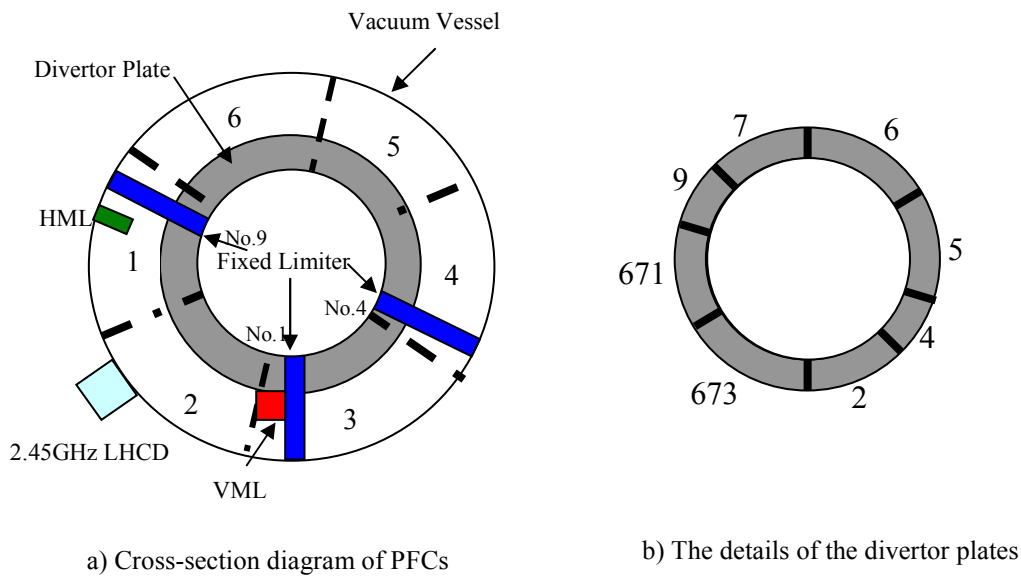

a) Cross-section diagram of PFCs     b) The details of the divertor plates

**Fig.1  Cross-section diagrams of PFCs**

The diagram a) shows the cross-section diagram of PFCs and the diagram b) shows the details of divertor plates. Numbering on the vacuum vessel and divertor plates show the branch of the cooling water channel.

limiters. A limiter installed on the top port of the vacuum vessel moves in the vertical direction, which has been called the vertical movable limiter (VML). The limiter head is composed of molybdenum brazed on the water cooled support made of stainless steel using thin copper foil attachment. As the result, thermal transport is better than other limiters. The moving distance could be varied by use of a bellows seal and a motor in the range of +32.5 mm and +80 mm, where the plus sign means the direction to the plasma side from the surface of the poloidal fixed ring limiter. The area of the plasma facing surface is $30 \times 160$ mm$^2$. The other movable limiter could shift its surface in the horizontal direction up to +23 mm which called the horizontal movable limiter (HML). A single probe is installed on the HML and the parameters of the SOL plasma have been estimated. The area of the plasma facing surface is $34 \times 74$ mm$^2$. This limiter is made of cupper covered with molybdenum. A molybdenum cover is fixed to the copper base by four screws, therefore thermal transport is not so good compared with the VML. The copper base attaches a water cooled support.

Two kinds of the microwave source in frequency of 2.45 and 8.2 GHz to excite the lower hybrid wave (LHW) and a microwave source in frequency of 170 GHz to inject the electron cyclotron wave (ECW) have been installed as the additional heating on TRIAM-1M. The microwave source in frequency of 2.45 GHz could produce up to 50kW in steady state and it mainly used in the experiments described in this paper. A source microwave excited by a crystal has been amplified by a klystron. The high power microwave transferred through the waveguide with rectangle cross-section to a launcher made of stainless steel for injecting the microwave to the plasma. A launching system is the $4 \times 1$ grill type waveguide array. The phase difference, $\Delta\Phi$, between adjacent waveguides can be changed by phase shifters [6]. On the 5 h operation, $\Delta\Phi=110$ degrees were used. Injection power is estimated by microwave power attenuated by the directional coupler in front of the inlet of the launcher. Attenuated microwave power was measured with corrected crystal diodes. Thus injected power can be measured during the discharge. For obtaining exact power absorbed to the plasma, it needs to estimate reflected power back to the launcher, uncoupling power, and power loss by propagation through the launcher. Reflected power back to the launcher is measured by the same way of that of injected power. The reflected microwave mainly excited by the coupling of the microwave and the plasma. Uncoupling microwave power was measured as leaked microwave power through the vacuum window made of sapphire. The correction of the power level is executed by the following way. Leak microwave power is measured at the window when the microwave is injected directly to the vacuum vessel. This gives the corrected relation of injected power to leaked power. Leaked power is measured in several discharges. These results are shown in Fig.2. As a result leaked power is estimated 7W every 1kW of injected power. Unfortunately the measurement of leaked power was not executed in the 5 h discharge. The estimated value of leaked power in 5 h discharge is shown in Fig.2. The microwave will also lose power via the dumping process of the microwave propagation in the waveguide. As the launcher is made of stainless steel, resistance of the surface of the waveguide is significantly high. Power loss on the launcher should be estimated to get net injected power.

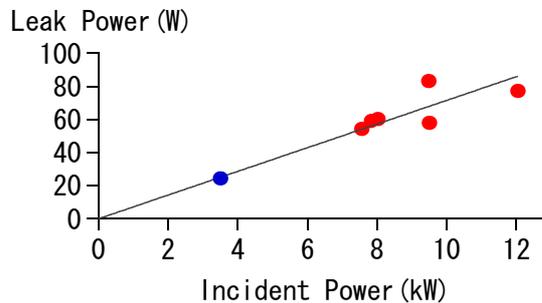

**Fig.2 Relation of incident power and leaked power**
Red dots are measured values and the blue dot means leaked power on 5 h discharge.

Cooling water systems are illustrated in Fig.3. Cooling water systems for the HML are independent from the other cooling system for PFCs. The main cooling water system has been used pure water produced by the ion-exchange resin method. The branch of the cooling water for 2.45GHz is connected to the main cooling water loop through only heat exchanger as shown in Fig.3. Main cooling water channels are divided 18 branches. Each part of the cooling water temperature is measured with

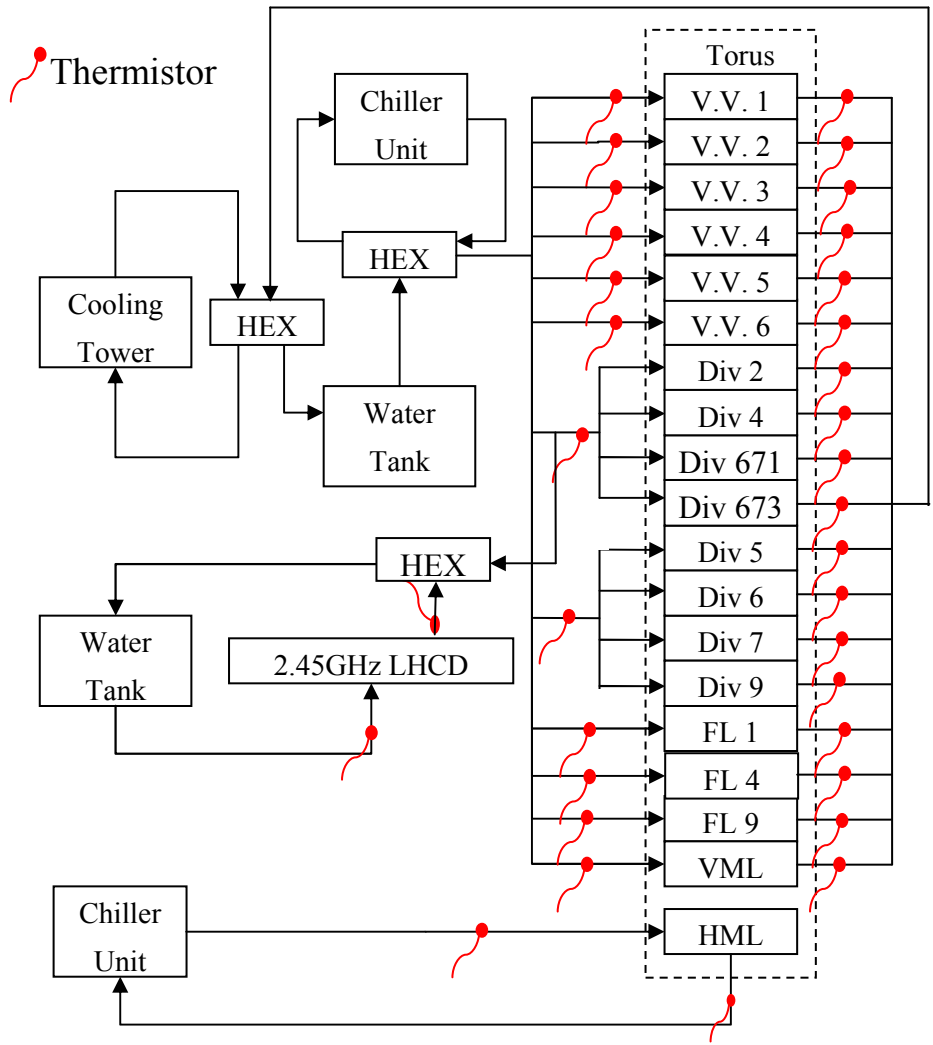

**Fig.3 Schematic diagram of the cooling water system**

The symbol of thermistor means the measurement point. All PFCs are measured with the inlet and out let temperature. Abbreviations V.V., Div, FL, HEX show the vacuum vessel, divertor plate, fixed limiter and heat exchanger respectively.

thermistors whose accuracy is comparable to 0.02 degree. The thermistors are thermal sensitive resistor and their resistance value changes significantly by the very small temperature change. The measurement spot is executed on the inlet and the outlet. Increment of water temperature between the inlet and the outlet, $\Delta T$, is necessary to estimate the heat load to each PFC. The measurement is implemented on 38 points (Fig.3). The data acquisition system is independent other system because the water temperature is required to measure for long time after the termination of the discharge until PFCs are cooled completely by water.

3. **Estimation of the heat load from the increment of water temperature**

In this section, the estimation of the heat load from the increment of cooling water temperature is described. The following equation is used to calculate the heat load from water temperature,

$$Q = \frac{C \cdot G \cdot \int \Delta T dt}{\tau_{dis}}, \qquad (1)$$

where $Q$ is the heat load [W], $C$ is specific heat of water [$J/g \cdot K$], $G$ is the flow rate of cooling water [g/sec], $\Delta T$ is the difference in temperature [K] and $\tau_{dis}$ is discharge duration [sec]. When the water temperature attains steady state, the simple equation without integration can be available.

$$Q = C \cdot G \cdot \Delta T. \qquad (2)$$

Equation (1) needs to measure water temperature until PFCs cooled completely. As the quite low temperature difference is difficult to measure, this makes the result undervalued. Poor cooling capability makes the low temperature difference for long time. As the result, the large error appears in the results. In fact, measuring until PFCs are cooled completely is difficult. On the other hand, the heat load can be estimated good accuracy when water temperature is steady state. Table 1 shows the flow rate of each PFC.

When water temperature of the inlet is varied, the temperature change of PFCs due to the change of the temperature of cooling water takes place. In this case, PFCs work as the heat sink or source. The heat load estimated in this situation includes the heat originated from the temperature change of PFCs. This leads to the large experimental error. Therefore the heat load to PFCs in 5 h discharge was estimated in the condition that the inlet water temperature is steady.

| Vacuum Vessel | Flow Rate(g/sec) | Divertor | Flow Rate(g/sec) | Limiter &LHCD | Flow Rate(g/sec) |
|---|---|---|---|---|---|
| No.1 | 138 | No.2 | 60 | FL No.1 | 74 |
| No.2 | 142 | No.4 | 71 | FL No.4 | 65 |
| No.3 | 140 | No.5 | 63 | FL No.9 | 43 |
| No.4 | 140 | No.6 | 68 | VML | 220 |
| No.5 | 137 | No.7 | 75 | HML | 217 |
| No.6 | 142 | No.9 | 72 | | |
| | | No.671 | 44 | Launcher | 367 |
| | | No.673 | 62 | | |

**Table 1 The flow rate of each PFC**

## 4. Experimental results

The typical wave form of 5 h discharge is shown as Fig.4. The fluctuation of water temperature of the VML and so on appears as shown in Fig.4. This is caused by the change of coupling between RF and plasma [7]. Here, the equation (2) can be used. The VML was inserted into plasma to remove the heat load effectively, therefore the last closed flux surface (LCFS) is decided by the VML. In this experiment, injected power which measured by corrected crystal diodes was 5710W and loss power at the launcher was estimated as 2220W by the increment of cooling water temperature and water flow rate. The estimated heat loads of each PFC are as the following.

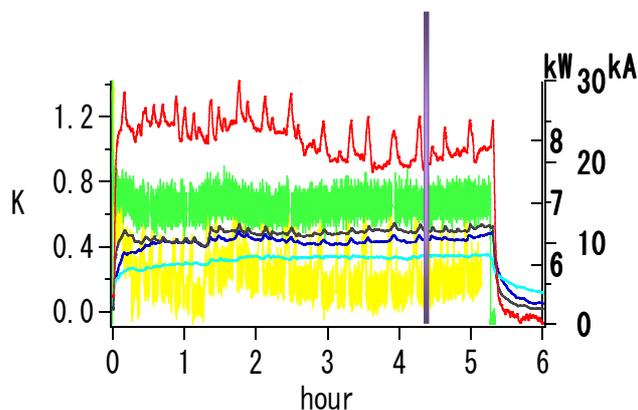

**Fig.4 Typical wave form of 5 h discharge**
Each line of color corresponds following;
Total injected power: Yellow
$I_p$: Green
Vacuum Vessel: Blue        Divertor: Light blue
Fixed limiter: Black       VML : Red
Power balance is calculated from data which included in the area of the purple marker.

| | |
|---|---|
| Fixed Limiters | 360W |
| Movable Limiter | 940W |
| Vacuum Vessel (include Divertor plates) | 1905W |

The heat load to limiters is mainly caused by the two processes, 1) the particle and the heat flux diffusing from the core plasma to the LCFS, and 2) the fast electron direct loss. The affect of the radiation to the heat load of limiters can be neglected, because the surface area of limiters is less than 2% of the total surface area of all PFC. Heat flux to limiters from the SOL plasma can be estimated by the electron temperature, $T_e$, and density, $n_e$ measured with a probe installed on the HML. To measure the profile of $T_e$ and $n_e$, the HLM is inserted in several discharges, which are maintained by LH power of 7.4 kW on corrected crystal diodes. Fig.5 shows the profile of $T_e$ and $n_e$ in the SOL region in this way. The head of the probe recedes 1 mm from the HML surface. The value of $T_e$ and $n_e$ do not change in the range of 5 mm and 7 mm despite the deep insert of the HML to the plasma. This indicates the head of the HML attaches to the LCFS. Thus the position of the LCFS can be decided in this way. The blue dotted curves show the fitted profile of electron density and temperature. The estimated heat flux to the HML derived from these data is 160W as referred in ref. [8]. The heat load to the VML is 590W, which include the heat load from the SOL plasma and the fast electron. The estimated value of the heat load from the SOL plasma is calculated as 350W by use of the profile of $T_e$ and $n_e$. The heat load to the other fixed limiter is mainly coming from the SOL plasma and it is 1100W. The total heat load from the SOL plasma corresponds to 1450W.

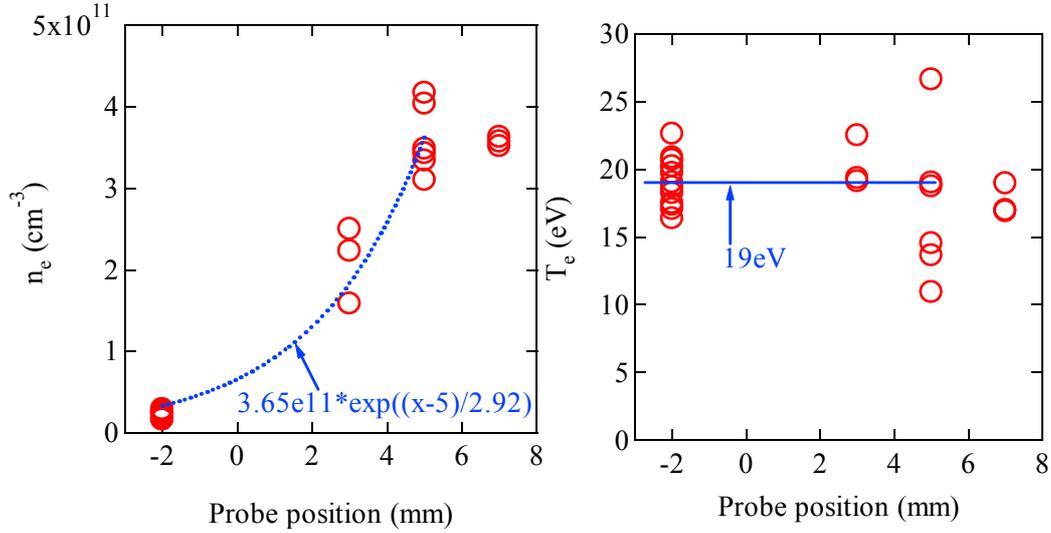

**Fig.5　Profile of $T_e$ and $n_e$**
Red circle means measured value and blue dotted line is fitted profile.

In the 5 h discharge, only VML is inserted into the plasma. Net injected RF power is 3.45 kW. Unfortunately plasma parameters in the SOL region are not measured in the 5 h discharge. We assume that the heat load from the SOL plasma is proportional to net injected RF power. This assumption is tolerable as shown in ref [9]. Under this assumption, the heat load from the SOL plasma may be estimated as 700W. As the result, the power of the fast electron loss is 760W in the 5 h discharge.

The fast electron accelerated by the microwave will be slowing down via the collision process. A part of fast electrons are sometimes scattered by the magnetic fluctuation and so on, and then they diffuse quickly to the LCFS. The orbit of the fast electron deviated from the flux surface. When the HML is attached to the LCFS, all of the fast electrons that located outside of the LCFS collides to the HML and the VML. As the result, power transferred by the direct loss of fast electrons can be measured as the heat load on the HML and the VML. The total heat load to the HML also includes from heat and the particle flux from the SOL plasma. Therefore the heat load from the SOL plasma should be subtracted from the total heat load to the HML in order to estimate the fast electron direct loss [10]. The total heat load to the VML and fixed limiters is 1460W. Therefore the direct loss power of the fast electron can be estimated as 1460-760=700W. The heat load to the vacuum vessel is caused by following processes, 1) radiation, 2) charge exchange (CX), 3) the particle and the heat flux from the core plasma via diffusion in the SOL region. The heat load via radiation will be deposited on the vacuum vessel and divertor plates, because radiation makes isotropic heat load and the vacuum vessel and divertor plates takes over all of area of PFCs. The heat deposition via CX process is similar to that of radiation but it is anisotropic because abundance of neutral particles is anisotropic. The cause of this inequality is attributable to the source of neutral particles. It is convenient that the distribution of Hα to estimate the localization of neutral particles. Fig.6 shows intensity of Hα [11]. 0 m of the abscissa axis corresponds to the location of the VML and a fixed limiter (No.1) of the toroidal direction and normalized by the intensity of this point. The location of intensity increasing exists to the limiter. It means that the limiters are source of neutral particle because they are exposed the SOL plasma.

Fig.7 shows the result of measurement of water temperature on the vacuum vessel. Fig. 7(a) shows the time evolutions of the increment of water temperature of No.1 and No.4 part of the vacuum vessels. The resemblance of these waveforms indicates that the heat load and cooling capability is almost same. On the other hand, the waveforms concerning No.3 and No.6 parts show the difference in the reached temperature increment. No.1 and No.4 parts have the same structure in the view of relative distance from a fixed limiter. On the other hand, No.3 part surrounds VML and a fixed limiter but No.6 vacuum vessel includes no limiter. Therefore, it is conceivable that this difference of the waveform is caused by the heat flux from CX process. The difference of the heat load between No.3 and No.6 parts are estimated 55 W. The direction of the No.3 vacuum vessel is 0 m to 0.6 m and that of No.6 part is 2.5 m to 3.1 m on Fig.6. When the difference of the heat load from CX process between No.3 and No.6 parts are estimated by the difference of the intensity of Hα in each part quantitatively under the assumption that the CX flux is proportional to the intensity of Hα. This consideration gives the total heat load of CX and it is calculated as 175W.

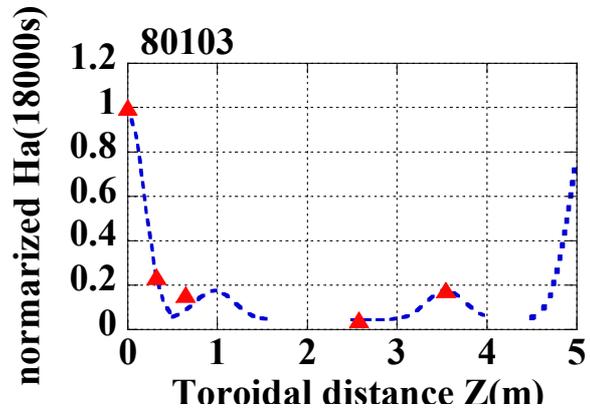

Fig.6 Distribution of Hα on toroidal direction
Red points are measured values and the blue dotted line is fitted line. It is seen that the rise on 0m, 1m, and 3.5m. There is VML and a fixed limiter on 0m, a fixed limiter on 1m and 3.5m..

## 5. Total power balance of the long time operation

From these results, power balance of the 5 h discharge can be estimated as below. Fig.8 shows the power balance of 5 h discharge. Power generated in 2.45GHz LHCD system (10260W) is lost in the launcher by damping (2220W) and reflected (4550W). Remaining power (5710W) is injected to the plasma. Although some power leaks (25W), and virtually power couples the fast electrons (3465W). Most of them give their energy by collisional slowing down on thermal electrons (2765W), and some of them collide to the VML as the direct loss of fast electron (700W). Energy to bulk plasma will be released via three processes, CX (175W), radiation (1730W) and diffusion (86W), respectively.

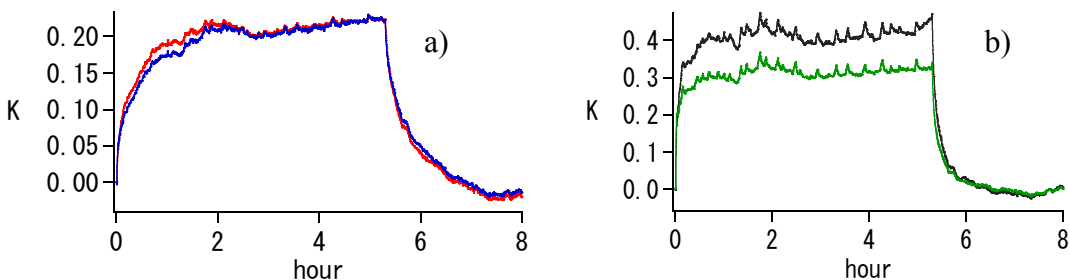

**Fig.7 Temperature change of cooling water on the vacuum vessel**
The graph a) shows the No.1 (red line) and No.4 (blue line) vacuum vessels and b) shows the No.3 (black line) and No.6 (green line). In the graph a), the results are coincident each other, but in the graph b), it is seen that there is the clear difference.

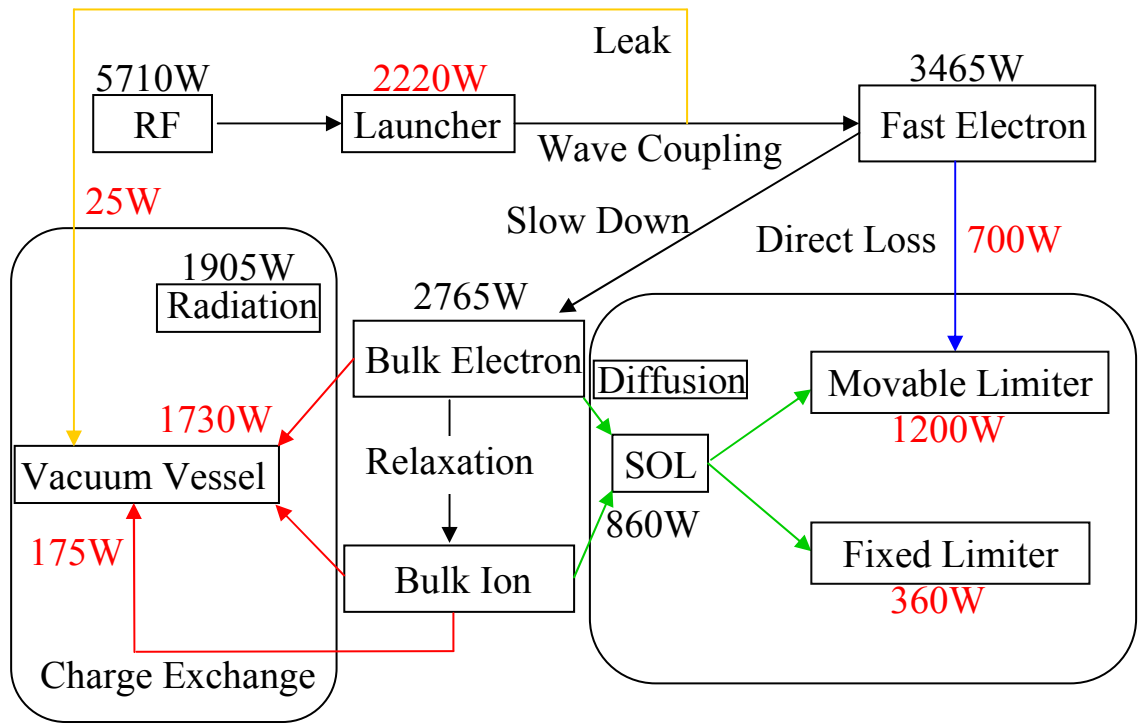

**Fig.8 Total power balance of the 5 h discharge**